\documentclass[11pt,final]{article} 
\usepackage{amssymb,amsthm,amsmath}
\usepackage{physics}
\usepackage{algorithm}
\usepackage{algorithmic}
\usepackage{graphicx}
\usepackage{float}
\usepackage{subcaption}
\usepackage{color}
\usepackage{xcolor}
\definecolor{pink}{rgb}{0.84, 0.04, 0.33}
\usepackage{booktabs}
\usepackage{longtable}
\usepackage{makecell}
\usepackage{array}
\usepackage{geometry}
\usepackage{marvosym}
\usepackage[numbers]{natbib}
\usepackage{url}

\title{COVID-19 in a social reinsurance framework: \\Forewarned is forearmed}
\author{\c{S}. \c{S}ahin${}^{1,3}$\footnote{\Letter\; Sule.Sahin@liverpool.ac.uk}, M.C. Boado-Penas${}^1$, C. Constantinescu${}^1$, J. Eisenberg${}^2$, \vspace{0.1cm}\\
	K. Henshaw${}^1$, M. Hu${}^1$, J. Wang${}^1$, W. Zhu${}^1$\vspace{0.6cm} \\
	University of Liverpool${}^1$, TU Wien${}^2$, Hacettepe University${}^3$}
\date{}

\begin{document}
\maketitle

\begin{abstract}
The crisis caused by COVID-19 revealed the global unpreparedness to handle the impact of a pandemic. In this paper, we present a statistical analysis of the data related to the COVID-19 outbreak in China, specifically the infection speed, death and fatality rates in Hubei province. By fitting distributions of these quantities we design a parametric reinsurance contract whose trigger and cap are based on the probability distributions of the infection speed, death and fatality rates. In particular, fitting the distribution for the infection speed and death rates we provide a measure of the effectiveness of a state's action during an epidemic, and propose a reinsurance contract as a supplement to a state's social insurance to alleviate financial costs. 
\vspace{6pt}
\vspace{.2in}
\\\textit{For far too long, our approach to pandemics has been one of panic and neglect: throwing money and resources at the problem when a serious outbreak occurs; then neglecting to fund preparedness when the news headlines move on. The result has been too many lives lost, too much damage to human livelihoods.} \citet{WB2017}.
\end{abstract}

\section{Introduction}

Pandemics, and in particular COVID-19, have significant impact on health systems, financial markets and vulnerable industries such as manufacturing, tourism and hospitality amongst others. Some governments have already announced economic measures to safeguard jobs and guarantee wages during the COVID-19 pandemic, but for the countries' economies this has a cost. For governments, planning and coordination are vital at the start of any epidemic to ensure that their health system is not overwhelmed and  to further alleviate the economic impact of the pandemic.\\

First reports of an outbreak of the (2019-nCov)-novel coronavirus-infected pneumonia (NCIP) were identified in December 2019 in the city of Wuhan, Hubei Province, China. Initially identified as pneumonia of unknown origin, confirmed cases were found to bear similarities to the severe acute respiratory syndrome coronavirus (SARS-CoV) of 2003 and the Middle East respiratory syndrome (MERS-CoV) of 2012, with the earliest instances of the virus linked to the Huanan Seafood Wholesale Market in Wuhan. Declared a pandemic by the World Health Organisation on 11 March 2020, the highly contagious virus has rapidly become a global concern, now present in more than 175 countries. \\

According to the World Health Organization (WHO), coronaviruses (CoV) are a large family of viruses that cause illnesses ranging from the common cold to more severe diseases such as SARS-CoV and MERS-CoV. The outbreak of SARS-CoV, \citet{DengPeng2020}, started in China in 2002 and was defeated by disease prevention and control systems. MERS-CoV was first reported in Saudi Arabia in 2012 and has since spread to several other countries. Although most of coronavirus infections, \citet{Huang2020clinical}, \citet{Sohrabi2020world} and \citet{zhu2020novel}, are not severe, more than 10,000 cumulative cases have been associated with SARS-CoV and MERS-CoV in the past two decades, with mortality rates of 10\% and 37\% respectively.\\

For governments there are two possible strategies to handle an epidemic: a) mitigation, which focuses on slowing the epidemic without interrupting transmission completely, i.e.\ reducing the peak healthcare demand while protecting vulnerable groups, and b) suppression which aims to reverse epidemic growth by reducing the number of cases to the minimum level. Mitigation involves isolating suspected cases, quarantining households with suspected cases and socially distancing the most vulnerable for around three months at the peak of the outbreak. Suppression includes the social distancing of the entire population with the added possibility of school and university closures. Mitigation, \citet{Ferguson2020}, was the strategy adopted by some US cities during the Spanish flu in 1918 and, in general, by the world in the 1957, 1968 and 2009 influenza pandemics. Suppression, while successful to date in China\footnote{Since the inital reporting of the outbreak in Wuhan, \citet{DengPeng2020}, \citet{Yang2020modified} and \citet{Sohrabi2020world}, China quickly entered a state of fighting against the new coronavirus based on their experience in the use of suppression policies during the SARS-CoV epidemic. Individual behaviour has been crucial in controlling the spread of COVID-19. As a result of the suppression measures taken by the government, on January 23rd 2020 Wuhan was quarantined and movement was restricted across Hubei province, affecting 50 million people. All public transportation was sealed off within China while outside of China aviation restrictions were applied and several countries initiated temperature and symptom screening protocols towards Chinese citizens. In Wuhan, two new hospitals were built in two weeks in early February to treat coronavirus patients.}
and South Korea, carries with it enormous social and economic costs which may significantly impact the well-being of society in the short and long run. Hence, there is a trade-off between minimising deaths from a pandemic and the economic impact of viral spread.\\

With the aim of ensuring an efficient healthcare response during a pandemic,  governments could engage in (re)insurance contracts at  state level to provide financial relief to both the state and the most vulnerable population stratum. According to \citet{WB2017}, the insurance industry can ensure rapid disbursement of funds to finance disaster response, and can create incentives for investing in risk mitigation and preparedness. Specifically for infectious diseases, the World Bank, and other partners, developed the Pandemic Emergency Financing Facility (PEF), an insurance vehicle designed to provide rapid disbursement of emergency finance. In practice, the World Bank collects the premiums and issues bonds and swaps to private investors, which can be seen as a type of catastrophe bond. So far the PEF has been widely criticised mainly due to the generous returns to investors and difficulty in accessing funding during the early stages of the disease outbreaks.\\

In general, the short-term mortality spikes caused by a pandemic have a tremendous impact on the world economy. Information about the expected severity and length of a possible outbreak is a corner stone for pricing a reinsurance contract. In recent years, in particular after the outbreak of SARS in 2002, scientists have warned about the possibility of a new pandemic. The  warning stated that most of the world is unprepared for such a challenge, see for instance \citet{WB2017}, since pandemics create unmanageable risks for life, travel and business insurance and ultimately the entire (re)insurance industry.\\

Apart from the devastating economic and political disruption at the state level, a pandemic has an additional effect of a micro-social-economic nature, bringing many families and individuals to the edge of poverty or even beyond. Unlike any other rare event such as a tsunami or an earthquake, a pandemic can last over relatively long periods, putting severe strain on a households' income through isolation restrictions. One of the roles of governments is to protect society against internal threats and to defend the country from any external threat that affects people's lives. COVID-19 represents a major threat to people's security. Allowing the state to purchase social reinsurance would limit the financial costs (i.e. income protection or any other type of unemployment benefit due to the effects of the pandemic) for the government to ensure the well-being of citizens in the case of a pandemic.\footnote{The contingencies associated with the reinsurance product would depend on the specifications of every contract.}
This is especially important for poor countries, where the financial resources are limited, or for countries with a high degree of income inequality. \\

The question arises of how a reinsurance company should quantify the risk from a pandemic and design a corresponding reinsurance contract. Although the reinsurance design and pricing challenges are not the core questions of the present paper, we highlight a path that links the already available data to the preparedness for a new pandemic at a national level. Pandemics call for fat-tailed distributions, see \citet{Taleb2020}. 
In \citet{WHO2018}, the authors suggest to calculate the expected loss in terms of the expected number of deaths, assigning a  ``statistical death'' to each country. Such a method could provide a reinsurer with a sense of  the expected losses to occur during a pandemic. However, without adequate measures and regulations in place, the losses might be almost unbounded, and therefore a cap for reinsurance coverage should be introduced. By estimating potential risks and in particular the necessary cap, reinsurance companies ``prepare for the last war,'' using the data available from comparable events in the past. There are several studies that estimate the basic reproductive numbers using epidemiological compartmental models. The reproductive number estimates the speed at which a disease is capable of spreading in a population, i.e. the number of secondary infected cases generated by one primary infected case.\footnote{For the interested reader see \citet{wu2020nowcasting}, \citet{Natsuko2020}, \citet{tang2020estimation}, \citet{li2020early}, \citet{yang2020mathematical} for estimates of reproductive numbers in China or \citet{wilkie} for estimates in the UK.}  \\

In this paper, we aim to provide an insight into an adequate reinsurance design with a coverage cap that provides an upper limit for reinsurance payments. Our analysis takes Hubei province in China as a benchmark due to the stringent measures implemented there since the start of the epidemic. If a particular country suffering the effect of a pandemic does not take sufficient measures to decrease the death rates and control the infection speed, the economic loss would be immense.  We adopt a data driven approach to calculate the economic loss caused by COVID-19 by exploring the probability distributions of the infection speed, death and fatality rates using the daily cross-sectional data obtained from 17 cities in Hubei province in China. Furthermore, we consider the parameter uncertainty in the probability distributions by treating the parameters as random variables over the course of the pandemic and investigating their distributions. Those parameters establish the coverage limit of our reinsurance contract and can be compared with the pre-specified ``barriers'' to determine the liability of the reinsurance company based on the relevant contract. By considering the infection speed, death and fatality rates, our approach not only provides a comprehensive perception of the losses caused by COVID-19 but also a measure of the effectiveness of the state's action during the pandemic in order to price an adequate reinsurance contract.\\

Following this introduction, the paper is structured as follows. In Section 2 we present some illustration of the data from China. Section 3 presents a descriptive analysis of the infection speed, death and fatality rates. In Section 4 we fit the distributions of the infection speed, death and fatality rates, and  propose a parametric reinsurance to serve as a social reinsurance contract for governments. We conclude in Section 5 and in Section 6, we define some of the distributions used in our data analysis as an appendix.

\section{Data} \label{Data}

As of 9 April 2020, the WHO reported 1,524,852 cumulative confirmed cases of COVID-19 in the world with 88,965 deaths and 332,989 recoveries. In China, the cumulative confirmed cases and number of deaths are 81,865 and 3,335 respectively. However the current number of infected cases is just 1,160, with this variable having decreased as of February 18th as a result of the measures imposed by the Chinese government in response to the coronavirus outbreak. As of 29 March 2020 the epicentre of the illness has been in Italy and Spain, with 95,262 and 85,043 active cases as of 9 April and a total number of 17,669 and 15,238 deaths respectively. In the USA the number of active cases is 397,472 with 14,797 deaths. In other countries such as Germany, France, Iran and the UK the number of confirmed cases is not negligible and has an exponential growth.\\

The R package, nCov2019, developed by \citet{nCov2019}, provides direct access to real-time epidemiological data of the outbreak. There are two kinds of data available from this package, real-time data and historical data. The real-time data, which contains current numbers of confirmed cases and deaths in geographical locations, are retrieved using API (application programming interface) calls to the Tencent SARS-COV-2 website \citep{tencent}. The Tencent website relies on official data obtained from the Chinese provincial health agencies, the China National Health Commission (CNHC), the World Health Organization (WHO) and public health agencies in other countries. \\

The historical data provided by the package, which forms the basis of our analysis, has three different sources. The first source is obtained directly from the CNHC.\footnote{CNHC holds the official historical statistics for the 34 provinces and special districts in China.} The second source is a non-governmental organisation Dingxiangyuan.\footnote{Dingxiangyuan has been continuously aggregating official data from provincial and city health agencies and the CNHC \cite{dxy}.} The third source is a public GitHub repository.
\footnote{This GitHub repository \cite{github} derives data from the literature \citet{Huang2020clinical} for December 1, 2019, to January 10, 2020, after which it relies on the Chinese news aggregator Toutiao API.} 
All three historical datasets are updated daily and are almost consistent with one other \citep{nCov2019_2}. We choose the third source since it provides the earliest data.
\begin{figure}[H]
	\begin{subfigure}{.4\textwidth}
		\includegraphics[scale=0.24]{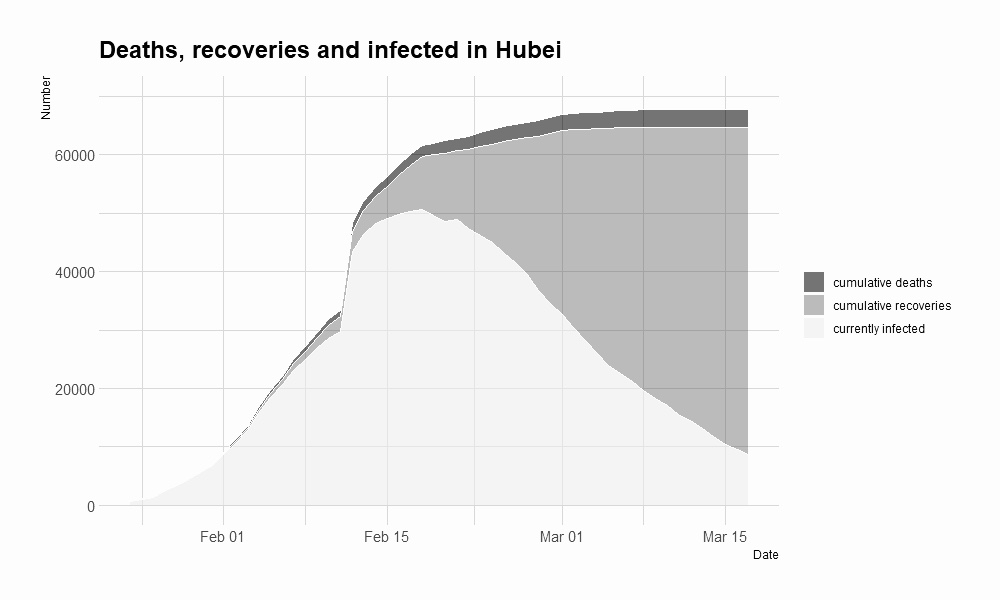}
		\caption{Number of cum. infected, recoveries and deaths in Hubei}\label{Hubei_data}
	\end{subfigure}%
	\hfill
	\begin{subfigure}{.48\textwidth}
		\includegraphics[scale=0.24]{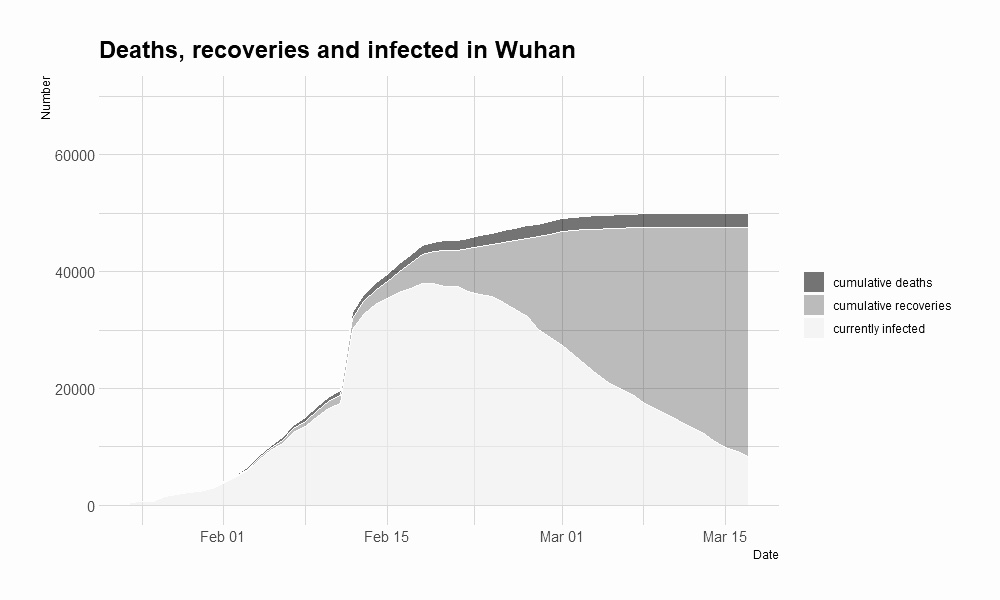}
		\caption{Number of cum.\ infected, recoveries and deaths in Wuhan}\label{Wuhan_data}
	\end{subfigure}
	\caption{Evolution of infected, recoveries and deaths in Hubei and Wuhan}
	\label{fig:evolution}
\end{figure}
Although not all data sources are official statistics, this kind of detailed data offers a unique opportunity to study the novel pathogen. In Figures \ref{Hubei_data} and \ref{Wuhan_data}, we observe that the epidemic was controlled by the end of February with almost no new cases. After this moment, an increase in the number of recoveries is seen in Hubei, particularly in Wuhan.\\

\begin{figure}[H]
	\begin{subfigure}{.5\textwidth}
		\includegraphics[scale=0.45]{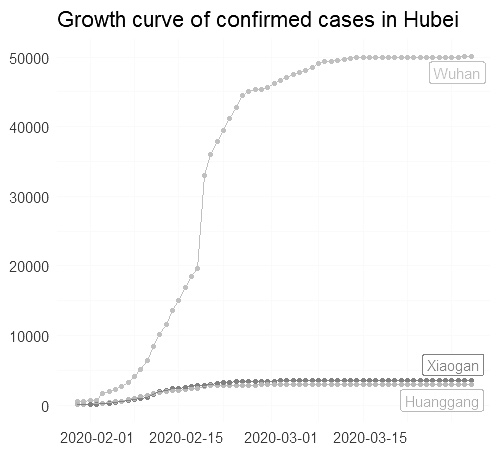}
		\caption{}
		\label{Hubei1}
	\end{subfigure}%
	\begin{subfigure}{.5\textwidth}
		\includegraphics[scale=0.45]{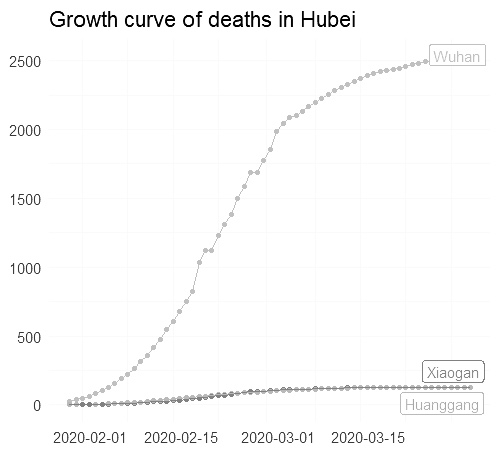}
		\caption{}
		\label{Hubei2}
	\end{subfigure}
	\caption{Growth curve data in Hubei}
	\label{fig:test}
\end{figure}

Hundreds of cities in China could even be considered as semi-independent outbreaks, as many are far from the epicentre and were effectively on lockdown from the end of January 2020. We particularly focused on the historical data in the epicentre, Hubei Province. The raw data contains the number of cumulative confirmed cases (Figure \ref{Hubei1}), cumulative deaths (Figure \ref{Hubei2}) and cumulative recovered cases. We use $\text{CC}_t$, $\text{CD}_t$ and $\text{CR}_t$ to denote the number of cumulative confirmed cases, cumulative deaths and cumulative recovered cases respectively at time $t$.

Our analysis will focus on three variables, infection speed, death rate and fatality rate. The infection speed for a city at time $t$, $v_t$ is defined by dividing the newly confirmed cases at time $t$ (the cumulative confirmed cases at time $t$ subtracted by the cumulative confirmed cases at time $t-1$) by the currently uninfected cases at time $t-1$ (the total population (TP) subtracted by the cumulative confirmed cases at time $t-1$)
$$
v_t=\frac{\text{CC}_t-\text{CC}_{t-1}}{\text{TP}-\text{CC}_{t-1}}.
$$

The death rate for a city at time $t$, $\Delta_t$  is defined by dividing the new deaths at time $t$ (the cumulative deaths at time $t$ subtracted by the cumulative deaths at time $t-1$) by the currently infected cases at time $t-1$ (the cumulative confirmed cases at time $t-1$ subtracted by the sum of cumulative deaths and recovered cases at time $t-1$)
$$
\Delta_t=\frac{\text{CD}_t-\text{CD}_{t-1}}{\text{CC}_{t-1}-\text{CD}_{t-1}-\text{CR}_{t-1}}.
$$
The fatality rate for a city at time $t$, $\psi_t$ is defined by dividing the cumulative deaths by the cumulative confirmed cases on that day
$$
\psi_t=\frac{\text{CD}_t}{\text{CC}_t}.
$$

Here we have two important assumptions. First, it is impossible to obtain daily updated population data, hence we use the official population data from the Hubei Statistical Yearbook 2019 published by the Hubei Provincial Bureau of Statistics \citep{Hubei}. Second, we assume recovered patients will be immune to the virus and so do not include recovered cases in the calculation of the infection speed. In fact, the second assumption will not affect the analysis significantly since the number of recovered cases is negligible in comparison to the total population.

\section{Descriptive Statistics}

Figure 3 shows the descriptive statistics for daily infection speed for the period 24 January - 17 March 2020, which were affected dramatically by the change in the way Chinese authorities accounted for confirmed cases as of 12 February. The mean and standard deviation graphs show sudden jumps on that specific day, however values are quite stable before and particularly afterwards. The median is much more robust to the outliers in the data than the mean and is not affected by the significant jump. Skewness and kurtosis values display similar patterns on different scales, both are mostly positive and consist of high values, an indication of a heavy tailed and right skewed distribution.  There are no recorded deaths in Wuhan on 21 February which affects the skewness and kurtosis values.

\begin{figure}[H]
	\centering
	\begin{minipage}[t]{0.45\textwidth}
		\vspace{2pt}
		\includegraphics[width=\textwidth]{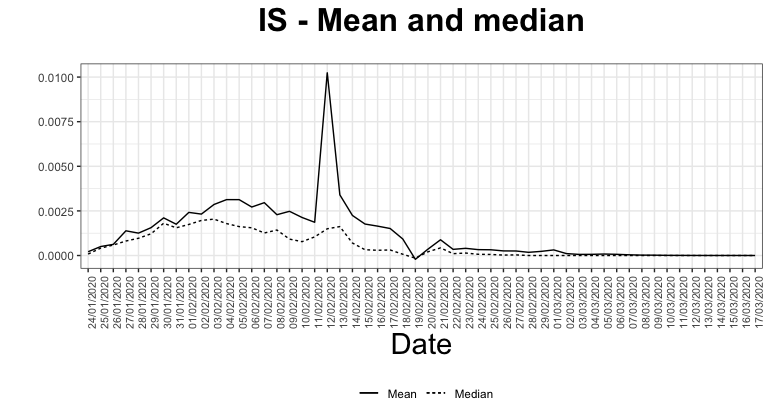}
		\subcaption{Daily time series graph for infection speed mean and median}
		\label{fig:ISMeanmed}
	\end{minipage}
	\hfill
	\begin{minipage}[t]{0.45\textwidth}
		\vspace{2pt}
		\includegraphics[width=\textwidth]{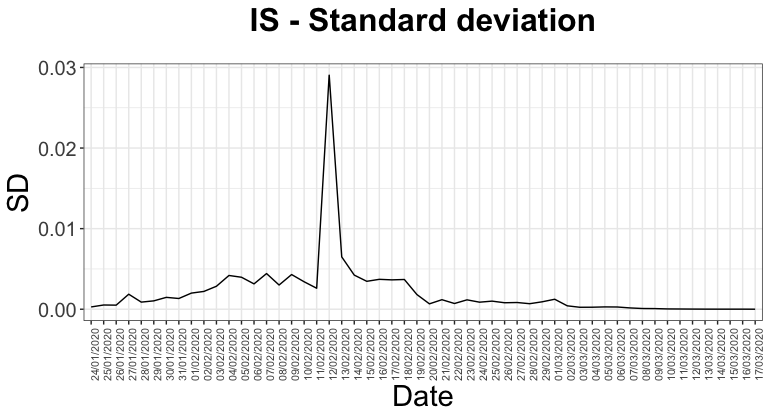}
		\subcaption{Daily time series graph for infection speed standard deviation}
		\label{fig:ISSD}
	\end{minipage}
	\hfill
	\begin{minipage}[t]{0.45\textwidth}
		\vspace{2pt}
		\includegraphics[width=\textwidth]{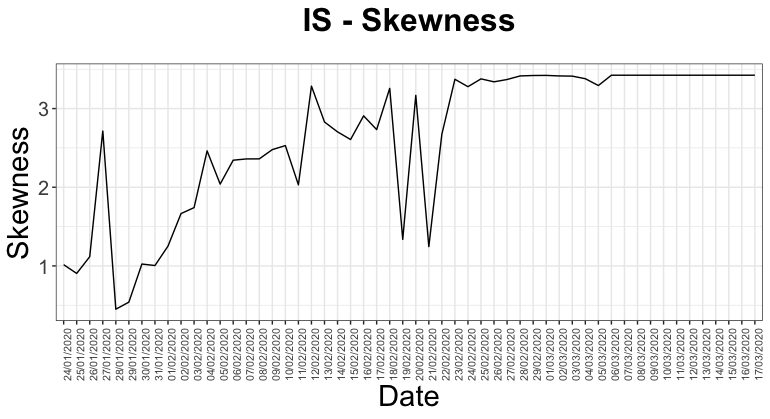}
		\subcaption{Daily time series graph for infection speed skewness}
		\label{fig:ISSkew}
	\end{minipage}\hfill
	\begin{minipage}[t]{0.45\textwidth}
		\vspace{2pt}
		\includegraphics[width=\textwidth]{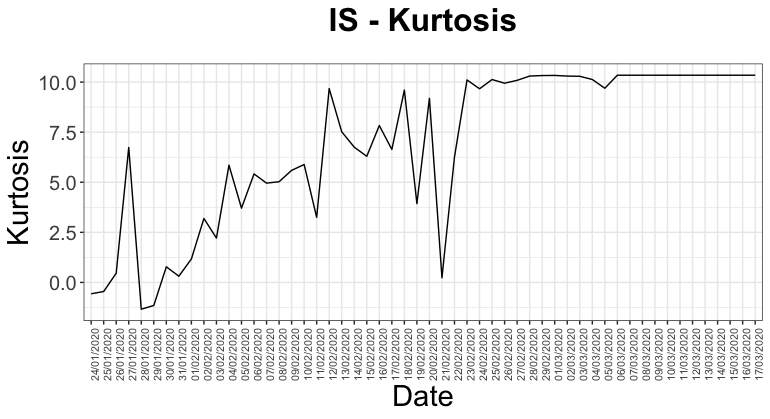}
		\subcaption{Daily time series graph for infection speed kurtosis}
		\label{fig:ISKurt}
	\end{minipage}
	\caption{Descriptive statistics for infection speed}
\end{figure}

\begin{figure}[H]
	\centering
	\begin{minipage}[t]{0.45\textwidth}
		\vspace{2pt}
		\includegraphics[width=\textwidth]{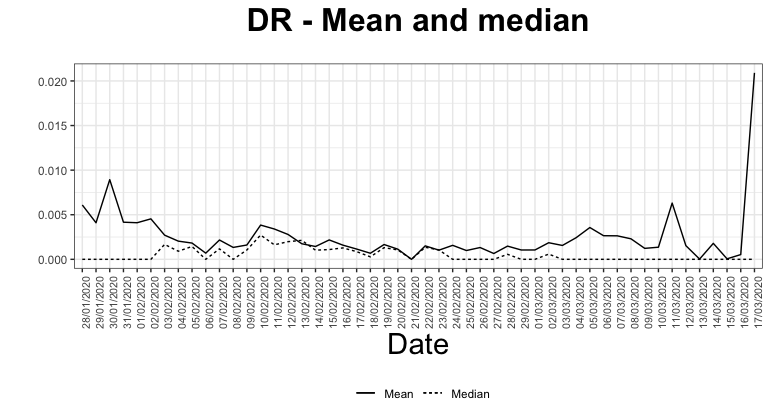}
		\subcaption{Daily time series graph for death rate mean and median}
		\label{fig:DRMeanmed}
	\end{minipage}
	\hfill
	\begin{minipage}[t]{0.45\textwidth}
		\vspace{2pt}
		\includegraphics[width=\textwidth]{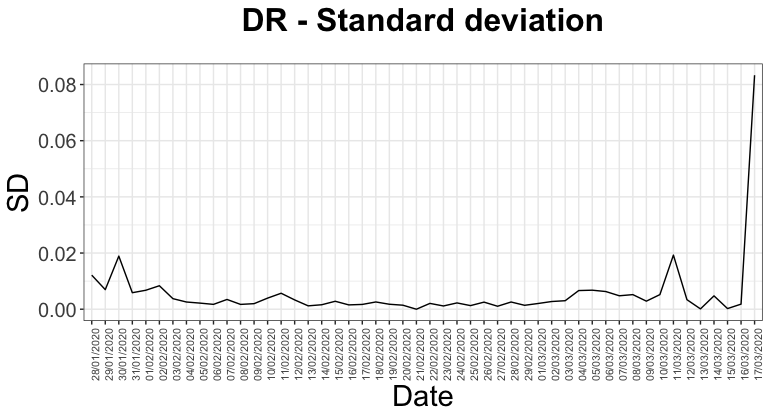}
		\subcaption{Daily time series graph for death rate standard deviation}
		\label{fig:DRSD}
	\end{minipage}
	\hfill
	\begin{minipage}[t]{0.45\textwidth}
		\vspace{2pt}
		\includegraphics[width=\textwidth]{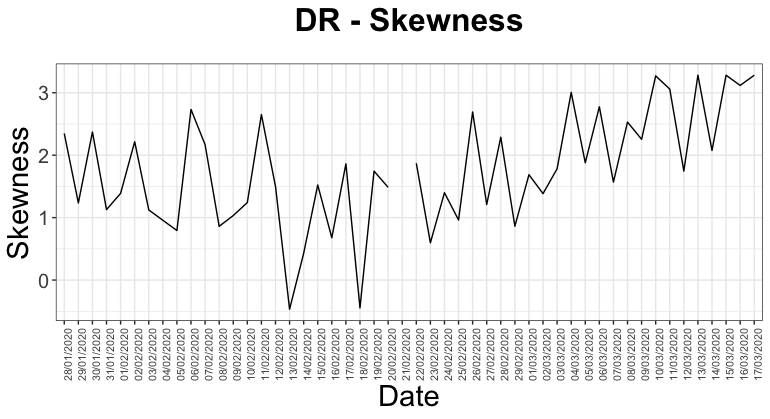}
		\subcaption{Daily time series graph for death rate skewness}
		\label{fig:DRSkew}
	\end{minipage}\hfill
	\begin{minipage}[t]{0.45\textwidth}
		\vspace{2pt}
		\includegraphics[width=\textwidth]{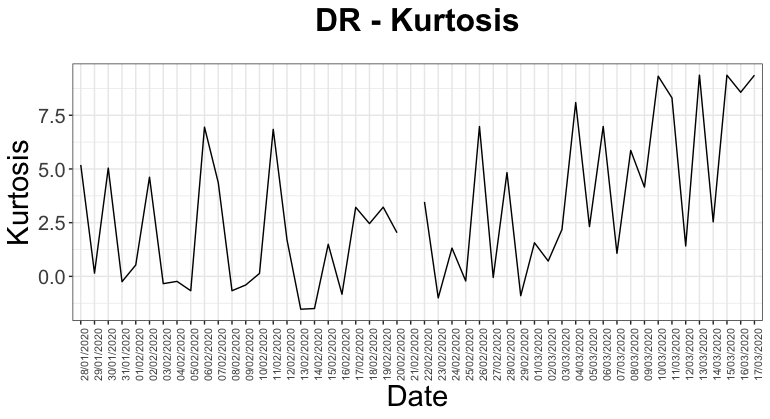}
		\subcaption{Daily time series graph for death rate kurtosis}
		\label{fig:DRKurt}
	\end{minipage}
	\caption{Descriptive statistics for death rate}
\end{figure}

As displayed in Figure \ref{fig:DRMeanmed}, the mean and median of the daily death rates, investigated over the period 28 January - 17 March 2020, are quite stable and present values which are close together from 6 February, the median values are however systematically lower than the mean values which may indicate a slight skewness to the right. The daily mean is around 0.0016 in the period following early February. Figure \ref{fig:DRSD} presents a low and stable standard deviation ranging from 0 to 0.002. At the end of the period the average death rate reaches a value of 0.0024, mainly due to the decrease in the number of new cases of infected people and thus a decrease in the number of deaths in the region. Skewness mainly fluctuates between 0 and 3 despite having a number of values both lower and higher, and the excess kurtosis is also extremely volatile and always positive, which may indicate a heavy tailed distribution for death rate time series data. 

\begin{figure}[H]
	\centering
	\begin{minipage}[t]{0.45\textwidth}
		\vspace{2pt}
		\includegraphics[width=\textwidth]{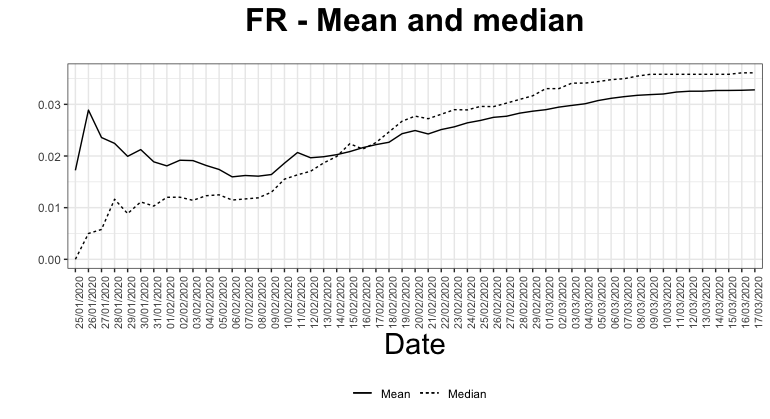}
		\subcaption{Daily time series graph for fatality ratio mean and median}
		\label{fig:FRMeanmed}
	\end{minipage}
	\hfill
	\begin{minipage}[t]{0.45\textwidth}
		\vspace{2pt}
		\includegraphics[width=\textwidth]{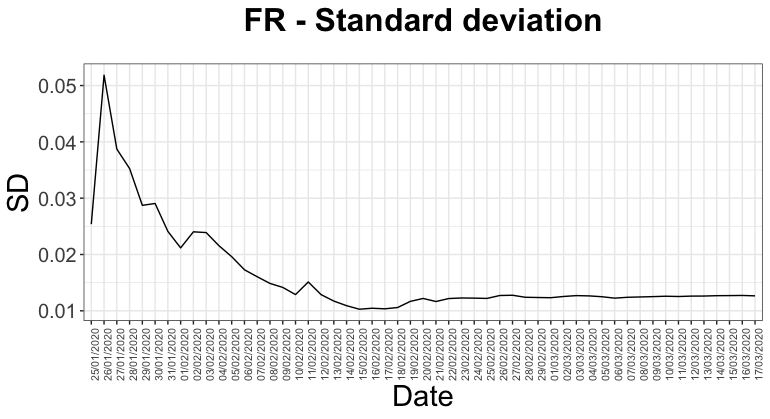}
		\subcaption{Daily time series graph for fatality ratio standard deviation}
		\label{fig:FRSD}
	\end{minipage}
	\hfill
	\begin{minipage}[t]{0.45\textwidth}
		\vspace{2pt}
		\includegraphics[width=\textwidth]{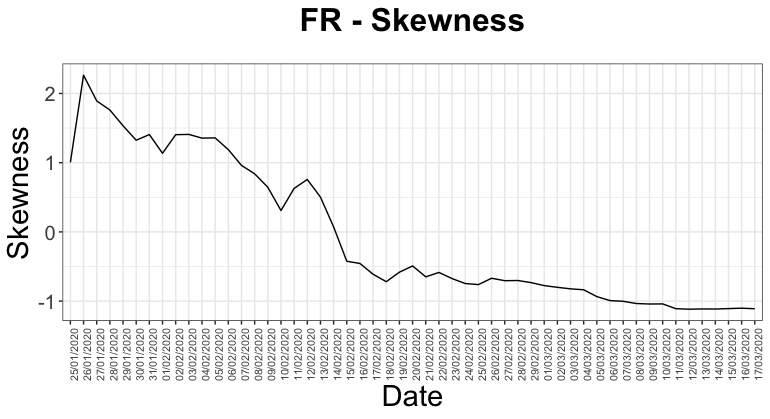}
		\subcaption{Daily time series graph for fatality ratio skewness}
		\label{fig:FRSkew}
	\end{minipage}
	\hfill
	\begin{minipage}[t]{0.45\textwidth}
		\vspace{2pt}
		\includegraphics[width=\textwidth]{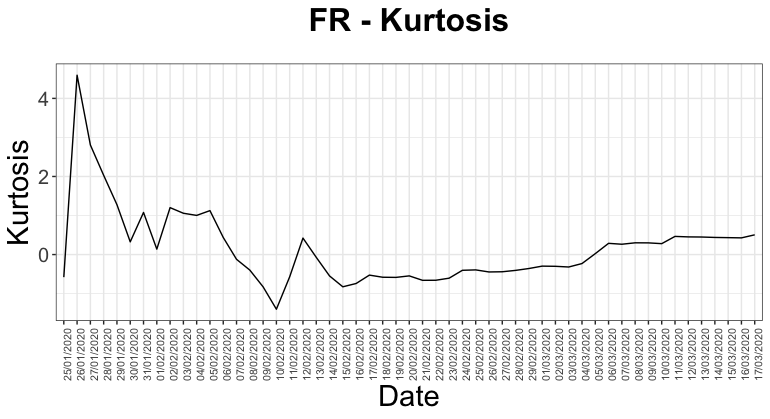}
		\subcaption{Daily time series graph for fatality ratio kurtosis}
		\label{fig:FRKurt}
	\end{minipage}
	\caption{Descriptive statistics for fatality ratio}
\end{figure}

Figure 5 presents the graphs of the descriptive statistics for daily fatality rates for the period 25 January - 17 March 2020. As seen in Figure \ref{fig:FRMeanmed}, the mean and median of the daily fatality rates have values which are distant from one another due to the zero deaths recorded in many cities during the early period of the outbreak. The close values of the mean and median after mid February indicates there are no outliers in the period under consideration and that the distributions of the daily fatality rates are not skewed significantly. The daily mean values of the fatality rates seem to stabilise around 3\% for the second half of the period analysed. We also observe in Figure \ref{fig:FRSD}, that the values of daily standard deviations are higher for the first half of the period while concentrated around 0.013 for the rest. The skewness, Figure \ref{fig:FRSkew}, taking positive values before 15 February and negative values afterwards indicates that the data has different distributional structures, although values are mostly between -1 and 1.  The daily excess kurtosis values, Figure \ref{fig:FRKurt}, are more volatile and present higher values before 15 February while displaying relatively small positive and negative values after this date, indicating a tail structure similar to the normal distribution.

\section{Methods employed and analysis of results} \label{Analysis}
Following the definition of the infection speed, death rate and fatality ratio, daily rates for each city in Hubei were calculated. A total of 61 distributions were fitted to daily speed data for which there are 41 days ranging from 24 January to 4 March 2020. Identifying the best three distributions based on Kolmogorov-Smirnov statistics for each day, the Generalised Extreme Value (GEV) distribution most frequently outperformed alternative distributions, fitting the data best in 28 cases. The same process was carried out for daily death rate, again fitting 61 distributions to 36 days of data from 28 January to 4 March 2020. The Normal distribution was the best fitting distribution in this case. For fatality ratio daily data, after fitting 61 distributions to 53 days of data from 25 January to 17 March 2020 and observing the associated Kolmogorov-Smirnov statistics for each day, the Johnson SB distribution was identified as the best fitting distribution. \\

\begin{figure}[H]
	\centering
	\begin{minipage}[t]{0.45\textwidth}
		\vspace{0pt}
		\includegraphics[width=\textwidth]{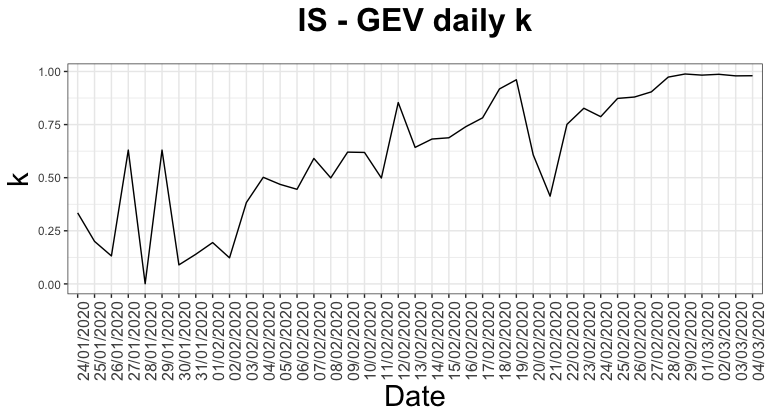}
		\subcaption{Daily time series graph for GEV parameter $k$}
		\label{fig:GEVk}
	\end{minipage}
	\hfill
	\begin{minipage}[t]{0.45\textwidth}
		\vspace{0pt}
		\includegraphics[width=\textwidth]{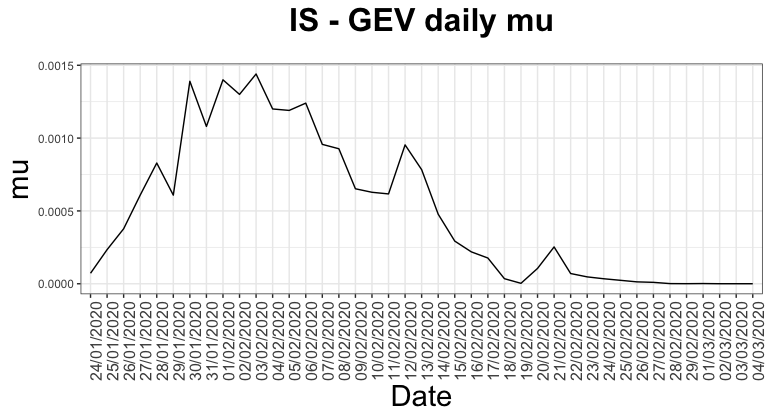}
		\subcaption{Daily time series graph for GEV parameter $\mu$}
		\label{fig:GEVmu}
	\end{minipage}
	\hfill
	\begin{minipage}[t]{0.45\textwidth}
		\vspace{2pt}
		\includegraphics[width=\textwidth]{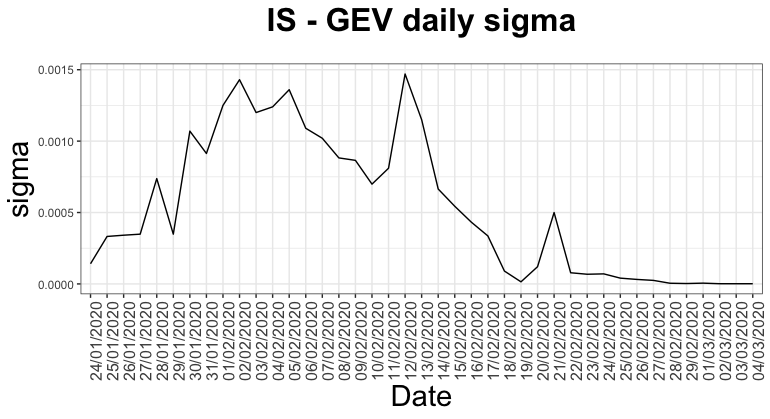}
		\subcaption{Daily time series graph for GEV parameter $\sigma$}
		\label{fig:GEVsigma}
	\end{minipage}
	\caption{Daily time series for GEV parameters}
	\label{timeseriesGEV}
\end{figure}

\begin{figure}[H]
	\centering
	\begin{minipage}[t]{0.45\textwidth}
		\vspace{0pt}
		\includegraphics[width=\textwidth]{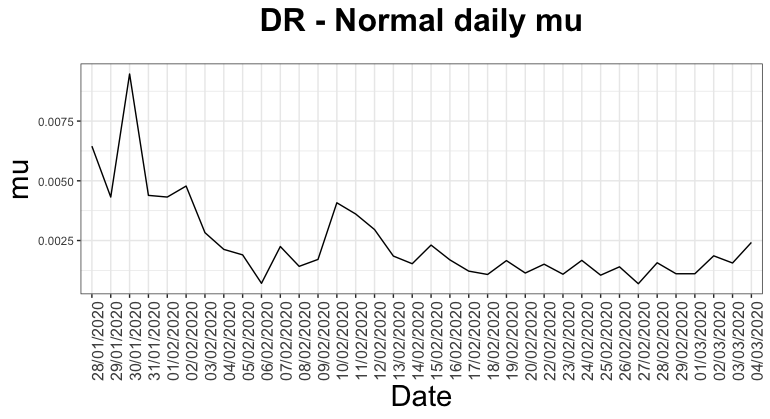}
		\subcaption{Daily time series graph for Normal parameter $\mu$}
		\label{fig:Normalmu}
	\end{minipage}
	\hfill
	\begin{minipage}[t]{0.45\textwidth}
		\vspace{0pt}
		\includegraphics[width=\textwidth]{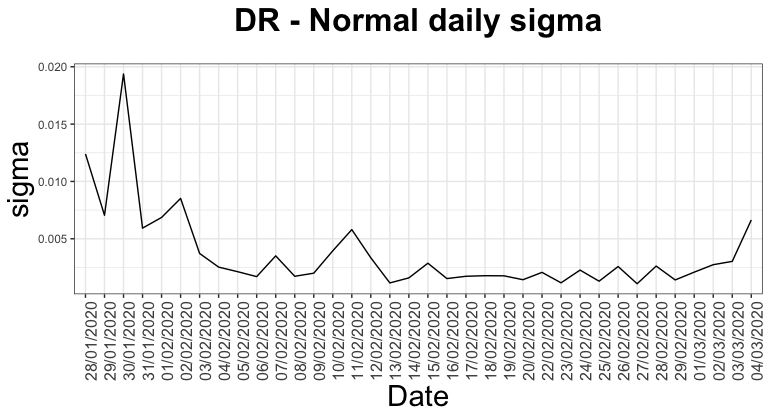}
		\subcaption{Daily time series graph for Normal parameter $\sigma$}
		\label{fig:Normalsigma}
	\end{minipage}
	\caption{Daily time series for Normal parameters}
	\label{timeseriesnormal}
\end{figure}

\begin{figure}[H]
	\centering
	\begin{minipage}[t]{0.45\textwidth}
		\vspace{0pt}
		\includegraphics[width=\textwidth]{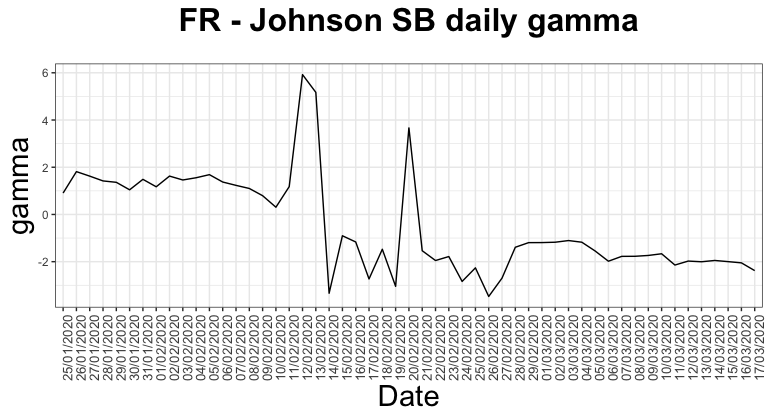}
		\subcaption{Daily time series graph for JSB parameter $\gamma$}
		\label{fig:JSBgamma}
	\end{minipage}
	\hfill
	\begin{minipage}[t]{0.45\textwidth}
		\vspace{0pt}
		\includegraphics[width=\textwidth]{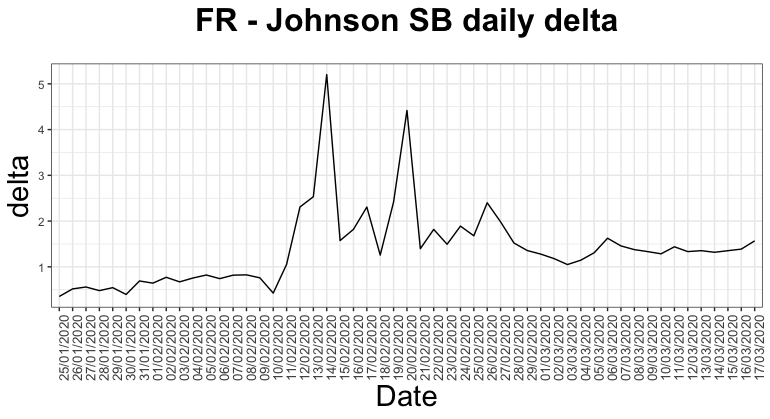}
		\subcaption{Daily time series graph for JSB parameter $\delta$}
		\label{fig:JSBdelta}
	\end{minipage}
	\hfill
	\begin{minipage}[t]{0.45\textwidth}
		\vspace{2pt}
		\includegraphics[width=\textwidth]{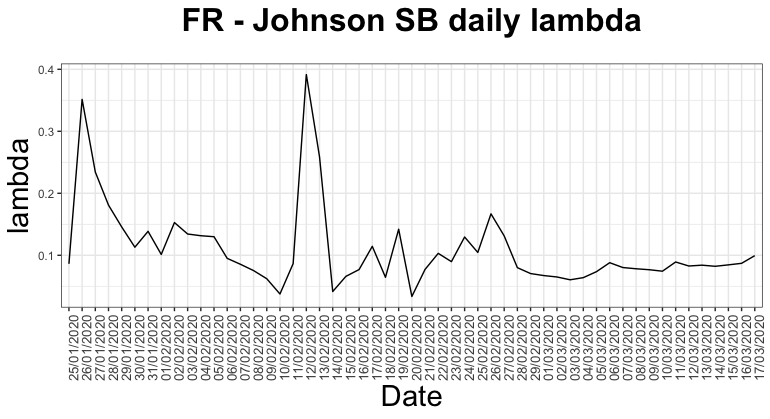}
		\subcaption{Daily time series graph for JSB parameter $\lambda$}
		\label{fig:JSBlambda}
	\end{minipage}
	\hfill
	\begin{minipage}[t]{0.45\textwidth}
		\vspace{2pt}
		\includegraphics[width=\textwidth]{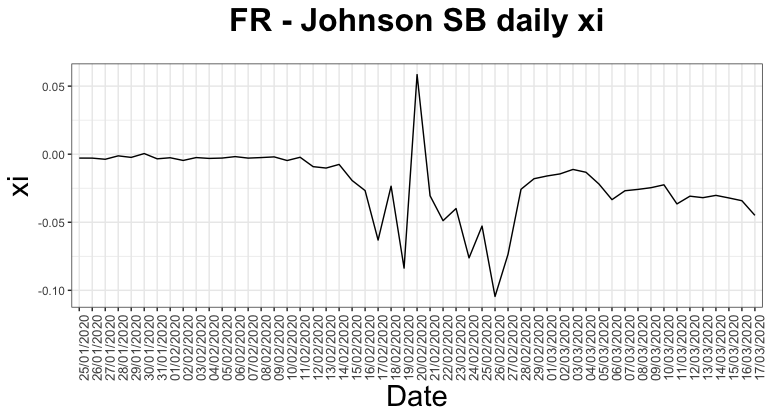}
		\subcaption{Daily time series graph for JSB parameter $\xi$}
		\label{fig:JSBxi}
	\end{minipage}
	\caption{Daily time series for Johnson SB parameters}
	\label{timeseriesJSB}
\end{figure}

Selecting the Generalised Extreme Value, Normal and Johnson SB distributions for speed, death and fatality data respectively, we fit only these distributions to each of the 41, 36 and 53 days. Parameters obtained were recorded and the best fitting distributions for each parameter identified. Further analyses of the model parameters including descriptive statistics and time series plots are detailed in Tables \ref{tableGEV}-\ref{JSBdescriptive} and Figures \ref{timeseriesGEV}-\ref{timeseriesJSB}, respectively. \\

For the three sets of data, infection speed, death rate and fatality ratio, we analyse the structure of the future losses caused by the coronavirus pandemic. Due to the lack of reliable data at the beginning of the pandemic in Wuhan, the diminishing number of confirmed cases and the occurrence of zero death rates in many cities by mid March, we choose periods with comparatively stable data flow for each data set, as described in Section \ref{Data}, and fit probability distributions to the daily Hubei province data. We use cross-sectional data by focusing on the infection speed, daily death rate and fatality ratio in 17 cities of Hubei, namely considering 17 observations per day. Once the best fitting distribution for each data set is decided, we consider the parameter uncertainty by examining the distribution of the parameters themselves. \\

Recall, our analysis aims to provide an insight into a social reinsurance design with a feasible deductible and a coverage cap, thus providing a band for the reinsurance payments. If a state affected by a pandemic does not take sufficient measures to decrease the death rate and infection speed, the economic loss may be unbounded. In particular, after a certain number of citizens have already been infected, the pandemic cannot be stopped. Therefore, a reinsurance company or more likely a conglomerate of reinsurance companies, would require a cap for the coverage of possible losses. Below we show how a cap can be defined by choosing a barrier for the parameters describing the future death rates or the infection speed, over a contractually pre-specified time interval. Based on our analysis, we will suggest a distribution and the corresponding parameters which should not exceed a certain interval indicating the cap for the cedent entity. \\

We consider the infection speed as it reflects in particular the activity of the government in preventing further outbreak of the infection, through for example, isolation measures, travel restrictions and the provision of face masks. As for daily speed data, a total of 61 distributions were fitted for the period beginning on 24 January and ending 4 March. Based on the Kolmogorov-Smirnov test statistics we identified the best three distributions for each day, selecting the most frequently fitted distribution to be the Generalised Extreme Value (GEV) distribution, and thus fit the GEV to each day. The probability density function of the GEV distribution is given by

\begin{equation*}
f(x) = \begin{cases} 
\frac{1}{\sigma}\exp(-(1+kz)^{-\frac{1}{k}})(1+kz)^{-1-\frac{1}{k}} & k\neq 0, \\
\frac{1}{\sigma}\exp(-z-\exp(-z)) & k=0, \\
\end{cases}
\end{equation*}
\\
\noindent where $k$ is the continuous shape parameter, $\sigma$ the continuous scale parameter ($\sigma>0$) and $\mu$ the continuous location parameter. The parameters for infection speed are plotted in Figure 6 and the descriptive statistics of the parameters presented in Table \ref{GEVdescriptive}.
The positive values of the shape parameter in Figure 6 indicate a heavy tailed distribution over the course of the pandemic while the location and scale parameters converge to zero rapidly during the second half of the period. This is consistent with the daily speed data for which values also become smaller and approach zero for most cities. The descriptive statistics in Table \ref{GEVdescriptive} show that $k$ has a symmetric distribution and the best fitting distribution for the shape parameter is the Johnson SB as given in Table \ref{tableGEV}. 

\begin{table}[H]
	\begin{center}
		\begin{tabular}{l|l|l|l|l|l|l|l|}
			\cline{2-8}
			&  Mean & SD & Median & Min & Max & Skewness & Kurtosis \\ \hline
			\multicolumn{1}{|l|}{$k$} & 0.6170 & 0.2920 & 0.6299 & 0.0008 & 0.9881 & -0.5015 & -0.7809  \\ \hline
			\multicolumn{1}{|l|}{$\mu$} & 0.0005 & 0.0005 & 0.0003 & 0.0000 & 0.0014 & 0.5914 & -1.1120 \\ \hline
			\multicolumn{1}{|l|}{$\sigma$} & 0.0005 & 0.0005 & 0.0003 & 0.0000 & 0.0015 & 0.4677 & -1.2250  \\ \hline
		\end{tabular}
	\end{center}
	\caption{Descriptive statistics for GEV parameters}
	\label{GEVdescriptive}
\end{table}

\begin{table}[H]
	\begin{center}
		\begin{tabular}{|c|c|c|cl}
			\cline{1-3}
			Parameter &  Distribution & Model Parameters &  &  \\ \cline{1-3}
			$k$ & Johnson SB distribution & \makecell{$\gamma=-0.5248$ \\ $\delta=0.72549$ \\ $\lambda=1.1917$ \\ $\xi=-0.13403$} &  &  \\ \cline{1-3}
			$\mu$ & Beta distribution & \makecell{$\alpha_1=0.30959$ \\ $\alpha_2=0.59326$ \\ $a=1.9019$e-07 \\ $b=0.00144$} &  &  \\ \cline{1-3}
			$\sigma$ & Beta distribution & \makecell{$\alpha_1=0.38219$ \\ $\alpha_2=0.67948$ \\ $a=1.3167$e-07 \\ $b=0.00147$} &  &  \\ \cline{1-3}
		\end{tabular}
		\caption{Best fitting distributions for parameters of the GEV distribution for daily infection speed data}
		\label{tableGEV}
	\end{center}
\end{table}

The location and scale parameters have negative excess kurtosis along with small skewness values, these are properties compatible with the Beta distribution whose probability density function is given by
\[f(x)=\frac{1}{B(\alpha_1,\alpha_2)}\frac{(x-a)^{\alpha_1-1}(b-x)^{\alpha_2-1}}{(b-a)^{\alpha_1+\alpha_2-1}},\]
where $B(\alpha_1,\alpha_2)$ is the Beta function
\[B(\alpha_1,\alpha_2)=\int^1_0t^{\alpha_1-1}(1-t)^{\alpha_2-1}dt.\]
Here $\alpha_1$ and $\alpha_2$ are continuous shape parameters ($\alpha_1>0$, $\alpha_2>0$) and $a$ and $b$ are continuous boundary parameters ($a<b$). The domain over which the distribution is defined is $a\leq x\leq b.$\\

The pdf of the Beta distribution has a U-shape with high density near the ends of the intervals $(0,0.0014)$ for $\mu$ and $(0,0.0015)$ for $\sigma$, and low density near $0.001$ for both parameters. Beta distributions with the given parameters put a lot of density on the extremes. 
The latter result is perfectly in line with the intuition behind the Beta distribution. The two extremes: the 
worst possible situation (highest $\mu$ and $\sigma$) and the best 
possible situation ($\mu$ and $\sigma$ equal zero) are the most likely 
events. Considering the famous baseball interpretation of the Beta 
distribution, we imagine the Chinese government as the baseball player, 
who has a success if the parameters $\mu$ and $\sigma$ are zero, i.e.\ 
the infection speed becomes zero, and a failure if the parameters attain 
the highest possible values in the support of the corresponding Beta 
distribution.\\

In addition to the intuitive explanation in support of selection of the distribution, the Beta distribution is widely used in data modelling since it is conjugate prior to the Bernoulli, 
binomial, negative binomial and geometric distributions in Bayesian 
inference. The computations in Bayesian inference may be very complex, however 
assuming a Beta distribution as a prior yields closed form formulas 
on the one hand and ensures the posterior distribution is a Beta 
distribution on the other.\\

The normal distribution is the best fitting distribution for the daily death rates, outperforming on 18 out of 36 days for the period 28 January to 4 March 2020. 
The domain over which the distribution is defined is $-\infty <x<+\infty$ and the probability density function of the normal distribution is given by
\[f(x)=\frac{\exp(-\frac{1}{2}(\frac{x-\mu}{\sigma})^2)}{\sigma\sqrt{2\pi}},\]
\noindent where $\sigma$ is the continuous scale parameter ($\sigma>0$) and $\mu$ the continuous location parameter. \\

The second and third most frequent best performing distributions for daily death rates are Johnson SB with 15 days and Generalised Pareto with 13 days. Graphs of the normal death rate parameters, location parameter $\mu$ and scale parameter $\sigma$, are shown in Figure 8. From early February both parameters have very small values and are fairly stable particularly for the last 15 days. The descriptive statistics in Table  \ref{Normaldescriptive} indicate heavy tailed distributions for both parameters, identified as the Burr distribution for $\mu$ and the Inverse Gaussian distribution for $\sigma$ with parameters presented in Table  \ref{tableNormal}.

\begin{table}[H]
	\begin{center}
		\begin{tabular}{l|l|l|l|l|l|l|l|}
			\cline{2-8}
			&  Mean & SD & Median & Min & Max & Skewness & Kurtosis \\ \hline
			\multicolumn{1}{|l|}{$\mu$} & 0.0024 & 0.0018 & 0.0017 & 0.0007 & 0.0095 & 2.2120 & 5.9997 \\ \hline
			\multicolumn{1}{|l|}{$\sigma$} & 0.0037 & 0.0037 & 0.0024 & 0.0011 & 0.0194 & 2.8292 & 9.5398  \\ \hline
		\end{tabular}
	\end{center}
	\caption{Descriptive statistics for Normal parameters}
	\label{Normaldescriptive}
\end{table}

\begin{table}[H]
	\begin{center}
		\begin{tabular}{|c|c|c|cl}
			\cline{1-3}
			Parameter &  Distribution & Model Parameters &  &  \\ \cline{1-3}
			$\mu$ & Burr distribution & \makecell{$k=0.35518$ \\ $\alpha=5.1224$ \\ $\beta=0.00125$} &  &  \\ \cline{1-3}
			$\sigma$ & Inverse Gaussian distribution & \makecell{$\lambda=0.00152$ \\ $\mu=0.00284$ \\ $\gamma=8.6910$e-04} &  &  \\ \cline{1-3}
		\end{tabular}
		\caption{Best fitting distributions for parameters of the Normal distribution for daily death rate data}
		\label{tableNormal}
	\end{center}
\end{table}

The death rates reflect in particular the effectiveness of actions concerning health care after being infected, such as building additional hospitals in record time and providing the necessary medical equipment. However, the parameters mean $\mu$ and variance $\sigma^2$ follow heavy tailed distributions. This means even if the death rates converge to zero relatively quickly, the probability for a large mean and variance remains high for quite a long time. The measures taken by a government may suppress the death rate, however the expectation of the rate might still increase, depending on the effectiveness of the measures. The variance being heavy-tailed coincides with the observations already made for COVID-19, that the severity of symptoms depends on the age of the infected person, generating in this way clusters in the set of the infected. Indeed, those older than 70 show more severe symptoms and have higher death rates than younger infected persons, \cite{Verity2020}.

We analysed daily fatality ratios for the 53 day period from 25 January to 17 March as the final data set. Since the fatality ratio is calculated by dividing cumulative deaths by cumulative confirmed cases, the values converge to specific rates due to the decrease in the death rate and the number of infected people. The Johnson SB distribution is selected as the best fitting distribution, outperforming for 18 days of the period under consideration. The probability density function of the Johnson SB distribution is given by 
\[
f(x)=\frac{\delta}{\lambda\sqrt{2\pi}z(1-z)}\exp\Big(-\frac{1}{2}\Big(\gamma+\delta\ln\Big(\frac{z}{1-z}\Big)\Big)^2\Big)\;,
\]
where $\gamma$ and $\delta$ are continuous shape parameters ($\delta>0$), $\lambda$ is the continuous scale parameter ($\lambda>0$) and $\xi$ the continuous location parameter. The domain over which the distribution is defined is $\xi \leq x\leq \xi+\lambda$. The parameter values plotted in Figure 8 are mostly volatile during the observation period, presenting several jumps particularly during the second half of February. The change in the method of counting the number of confirmed cases on 12 February affects calculation of fatality rates, as the number of cumulative confirmed cases is used directly in the calculation. All parameters reflect drastic change, with the jumps also influencing the distribution of the parameters. The descriptive statistics presented in Table \ref{JSBdescriptive} indicate that the positive shape parameter $\delta$, and the scale parameter $\lambda$, have heavy tailed distributions. The best fitting distributions for these parameters are given in Table \ref{tableJSB}, which identifies the Weibull distribution for $\delta$ and the GEV for $\lambda$, while the location parameter $\xi$, has Hypersecant distribution and the shape parameter $\gamma$, Generalised Pareto distribution.

\begin{table}[H]
	\begin{center}
		\begin{tabular}{l|l|l|l|l|l|l|l|}
			\cline{2-8}
			&  Mean & SD & Median & Min & Max & Skewness & Kurtosis \\ \hline
			\multicolumn{1}{|l|}{$\gamma$} & -0.4433 & 2.0951 & -1.1911 & -3.4736 & 5.9231 & 0.9555 & 0.7389 \\ \hline
			\multicolumn{1}{|l|}{$\delta$} & 1.3961 & 0.8823 & 1.3330 & 0.3523 & 5.2017 & 2.3186 & 7.7067  \\ \hline
			\multicolumn{1}{|l|}{$\lambda$} & 0.1098 & 0.0679 & 0.0866 & 0.0336 & 0.3914 & 2.5645 & 7.6142 \\ \hline
			\multicolumn{1}{|l|}{$\xi$} & -0.0218 & 0.0261 & -0.0180 & -0.1045 & 0.0585 & -0.6753 & 2.6873  \\ \hline
		\end{tabular}
	\end{center}
	\caption{Descriptive statistics for Johnson SB parameters}
	\label{JSBdescriptive}
\end{table}

\begin{table}[H]
	\begin{center}
		\begin{tabular}{|c|c|c|cl}
			\cline{1-3}
			Parameter &  Distribution & Model Parameters &  &  \\ \cline{1-3}
			$\gamma$ & Generalised Pareto distribution & \makecell{$k=-0.36417$ \\ $\sigma=3.6692$ \\ $\mu=-3.133$} &  &  \\ \cline{1-3}
			$\delta$ & Weibull distribution & \makecell{$\alpha=1.2597$ \\ $\beta=1.133$ \\ $\gamma=0.34228$} &  &  \\ \cline{1-3}
			$\lambda$ & Generalised Extreme Value distribution & \makecell{$k=0.35458$ \\ $\sigma=0.02823$ \\ $\mu=0.07851$} &  &  \\ \cline{1-3}
			$\xi$ & Hypersecant distribution & \makecell{$\mu=-0.02176$ \\ $\sigma=0.02611$} &  &  \\ \cline{1-3}
		\end{tabular}
		\caption{Best fitting distributions for parameters of the Johnson SB distribution for daily fatality ratio data}
		\label{tableJSB}
	\end{center}
\end{table}

Based on our analysis, we propose two possibilities which can be extended by different options for a reinsurance contract.
The first suggestion is to take as a basis a GEV distribution for the infection speed, a normal distribution for the death rate and to choose $\hat \mu_s$ and $\hat\mu_d$ as coverage limit parameters for the infection speed and death rates respectively. Observing the data during a new 
pandemic, the reinsurance company will compare the location parameter obtained under the assumption of a GEV distribution with 
the given ``barrier'' $\hat \mu_s$ and the mean of the death rate under assumption of a normal distribution with the barrier $\hat\mu_d$. If the 
obtained parameters exceed (here we can choose between different definitions of exceeding a barrier in a multidimensional setup) the barriers, the reinsurance company will 
reinstate just the amount assigned to the fixed vector $(\hat \mu_s,\hat\mu_d)$. In this way we take into account both features of government's actions: the health care of the already infected and the measures preventing further outbreak. \\

The second possibility would be to use a one-dimensional cap -- the fatality rate. Fatality rates combine the death rate and infection speed in a natural way that cannot be observed directly.
Similarly to the above described procedures, we assume a Johnson SB distribution for the fatality rates, fix the limit location parameter to be $\hat \xi$ and observe the location parameter $\xi$ of the new pandemic. If the 
obtained parameter exceeds the barrier $\hat \xi$, the reinsurance company will 
reinstate just the amount assigned to $\hat \xi$.
We assume that the assignment procedures will 
follow a predefined method where a statistical death is linked to an 
average loss connected to a specific country, region or even population 
stratum, see for instance \citet{WHO2018}.\\

Hence, we are proposing a type of a parametric reinsurance, where not (only) the trigger but the cap is linked to an event parameter, a vector containing location parameters of the death rate and infection speed or of the fatality ratio. Of course, one can also extend the number or the character of triggers and/or caps by adding for example the scales to the event parameter vector. Once the reinsurance mechanism is triggered, payments will be made regardless of the actual physical loss sustained, a cap is applied if the parameter threshold is reached or exceeded.\\

Big reinsurance companies like for example Swiss Re use parametric products for events such as earthquakes, tropical cyclones, or floods, where the parameters are magnitude, wind speed or precipitation respectively, see \cite{swissre}.
\\
We do not propose to replace traditional reinsurance products by a parametric reinsurance. We suggest to fill an existing gap, left by traditional insurance, see \citet{WHO2018}, where the insured, in our case a state, has no control over the emergence and initial spread of a highly contagious and deadly infection.
However, a government can take measures against the quick spread and high mortality once a potential danger has been detected. The parametric (re)insurance design we suggest will measure the effectiveness of government's actions and indemnify the losses on this basis, preventing the bankruptcy of the reinsurer and providing the necessary help to the vulnerable.\\

The level of testing, accuracy of the diagnosis and efficiency of methods for recording confirmed, recovered and death cases are of crucial importance to the collected data. Although we are aware the robustness of our analysis are subject to the limitations imposed by the data, the fitted distributions and the distributions of the parameters provide crucial information regarding the evolution of the pandemic and the economic loss it might cause.  Focusing on the GEV infection speed parameters, one can easily infer that the pandemic in China has been taken under control, with the time series graphs of the location and scale parameters heading to zero by mid February. This shows that measures such as lockdown, isolation, face masks, and so on, taken by the Chinese government decreased the number of newly infected people significantly and thus the infection speed. On the other hand, being distributed normally, the death rates are forced to lower as of early February, exhibiting some volatility during the course of the pandemic. The actions taken by governments do not only determine the associated loss  but also the duration of the pandemic and this can be observed in the parameter values recorded over time. Similar comments are valid for the fatality ratios as the parameters of the fitted Johnson SB distribution start to flatten from early March, the effect of government interventions is visible in the ratios, which are based on cumulative deaths and cumulative confirmed cases, over a longer period.

\section{Conclusion}

Since mid-February in China, where the novel coronavirus was initially reported, the number of infected people has steadily declined thanks to the extreme measures the government took to contain the virus. Currently the only new cases in China are cases brought in from abroad.\\

As of 9 April 2020, more than 184 countries have reported cases of COVID-19. In particular, the WHO reported 1,524,852 cases in the world including 88,965 deaths. In terms of financial impact, the shock to the global economy from COVID-19 has been faster and more severe than the 2008 global financial crisis and even the Great Depression. For governments developing capacities for mitigating the effects of a pandemic,  health security is a priority.  Activities such as planning and coordination are vital at the start of an epidemic or a pandemic.  In response to the COVID-19 outbreak, some governments have launched unprecedented public health and economic responses. In recent weeks, more than 80 poor and middle-income countries have sought financial help from the International Monetary Fund as they struggle to cope with the economic fallout from the COVID-19 pandemic. We hope our analysis will help inform such decisions.\\

The social reinsurance design provided in this paper can be considered a supplement to the social insurance of the state and it could be a solution to alleviate the financial costs for governments in the event of a pandemic. It is a type of parametric reinsurance whose trigger and cap are based on the probability distributions of the infection speed, death and fatality rates. As this social resinsurance product design involves a cap, it forces governments to take the measures needed to fight against the pandemic. Based on the methodology described, it would be interesting to analyse how effective the measures taken by other countries are in terms of infection speed, mortality rate and fatality ratio. Another direction of research would be to calculate the expected shortfall of the loss depending on the measures taken by governments.\\

We would like to close with ``Recommendation 10'' of the \citet{WB2017}: ``To reinforce incentives for national governments to invest in preparedness, the IMF and World Bank should work to facilitate the incorporation of the economic risks of infectious disease outbreaks into macroeconomic and market assessments, including: (i) inclusion into assessments where such risks are macro-critical; (ii) encouraging the development of academic and private sector indices and maps of intrinsic risk, preparedness and economic vulnerability.'' Maybe we will be forearmed for the next one.

\section{Relevant distributions}

In \citet{Taleb2020}, analysis of the history of pandemics suggests a fat-tailed distribution for the number of  deaths, which could always have a Generalized Pareto Distribution (GPD) as a tail approximation.
The distributions we identified as the best fitting distributions for the infection speed, death rate and  fatality ratio are mostly fat-tailed, and the ones not defined in the paper, are defined below.\\

\noindent \underline{Burr Distribution:}

The Burr distribution has parameters $k$, $\alpha$, $\beta$ and $\gamma$, where $k$ and $\alpha$ are continuous shape parameters ($k>0$, $\alpha>0$), $\beta$ is the continuous scale parameter ($\beta>0$) and $\gamma$ the continuous location parameter ($\gamma \equiv 0$ yields the three-parameter Burr distribution). The domain over which the distribution is defined is
\[\gamma \leq x < +\infty.\]
The probability density function of the four-parameter Burr distribution is given by
\[f(x)=\frac{\alpha k(\frac{x-\gamma}{\beta})^{\alpha-1}}{\beta(1+(\frac{x-\gamma}{\beta})^{\alpha})^{k+1}}.\]\\

\noindent \underline{Inverse Gaussian Distribution:}

The Inverse Gaussian distribution has parameters $\lambda$, $\mu$ and $\gamma$, where $\lambda$ and $\mu$ are continuous parameters ($\lambda>0$, $\mu>0$) and $\gamma$ is the continuous location parameter ($\gamma \equiv 0$ yields the two-parameter Inverse Gaussian distribution). The domain over which the distribution is defined is
\[\gamma < x< +\infty.\]
The probability density function of the three-parameter Inverse Gaussian distribution is given by
\[f(x)=\sqrt{\frac{\gamma}{2\pi(x-\gamma)^3}}\exp\Big(-\frac{\lambda(x-\gamma-\mu)^2}{2\mu^2(x-\gamma)}\Big).\]\\

\noindent \underline{Generalized Pareto Distribution:}

The Generalized Pareto distribution has parameters $k, \sigma$ and $\mu$, where $k$ is the continuous shape parameter, $\sigma$ the continuous scale parameter ($\sigma>0$) and $\mu$ the continuous location parameter. The domain over which the distribution is defined is
\begin{align*}
\mu \leq x <+\infty \ \ &\text{for} \ \ k\geq 0, \\
\mu \leq x \leq \mu-\frac{\sigma}{k} \ \ &\text{for} \ \ k<0.
\end{align*}
The probability density function of the Generalized Pareto distribution is given by
\begin{equation*}
f(x) = \begin{cases} 
\frac{1}{\sigma}(1+k\frac{(x-\mu)}{\sigma})^{-1-\frac{1}{k}} & k\neq 0, \\
\frac{1}{\sigma}\exp(-\frac{(x-\mu)}{\sigma}) & k=0. \\
\end{cases}
\end{equation*}\\

\noindent \underline{Weibull Distribution:}

The Weibull distribution has parameters $\alpha$, $\beta$ and $\gamma$, where $\alpha$ is the continuous shape parameter ($\alpha>0$), $\beta$ the continuous scale parameter ($\beta>0$) and $\gamma$ the continuous location parameter ($\gamma \equiv 0$ yields the two-parameter Weibull distribution). The domain over which the distribution is defined is
\[\gamma \leq x< +\infty.\]
The probability density function of the three-parameter Weibull distribution is given by
\[f(x)=\frac{\alpha}{\beta}\Big(\frac{x-\gamma}{\beta}\Big)^{\alpha-1}\exp\Big(-\Big(\frac{x-\gamma}{\beta}\Big)^{\alpha}\Big).\]\\

\noindent \underline{Hyperbolic Secant Distribution:}

The Hyperbolic Secant distribution has parameters $\sigma$ and $\mu$, where $\sigma$ is the continuous scale parameter ($\sigma>0$) and $\mu$ the continuous location parameter. The domain over which the distribution is defined is
\[-\infty < x< +\infty.\]
The probability density function of the Hyperbolic Secant distribution is given by

\[ f(x)=\frac{\sec{(\frac{\pi(x-\mu)}{2\sigma})}}{2\sigma}.\]

\bibliography{Corona1}

\end{document}